\documentclass{sig-alternate-05-2015}
\usepackage{hyperref}
\usepackage{cite}
\usepackage{url}
\usepackage[tight,footnotesize]{subfigure}
\usepackage{balance}
\usepackage{multirow}
\usepackage{fancybox}
\usepackage{mathrsfs}
\usepackage{wrapfig}
\usepackage{graphicx}
\usepackage{amsmath}
\usepackage{mathtools}
\usepackage[mathscr]{euscript}
\usepackage{algorithm}
\usepackage{algorithmic}

\usepackage{amsthm}
\theoremstyle{definition}

\newtheorem{example}{Example}[section]
\theoremstyle{remark}

\graphicspath{{figures/}}
\usepackage{ctable}
\usepackage{bbm}
\usepackage{verbatim}
\usepackage{comment}
\newcommand{\leqnos}{\let\@eqnnum\l@eqnnum}
\newcommand{\reqnos}{\let\@eqnnum\r@eqnnum}

\begin{document}

% Copyright
%\setcopyright{acmcopyright}
%\setcopyright{acmlicensed}
%\setcopyright{rightsretained}
%\setcopyright{usgov}
%\setcopyright{usgovmixed}
%\setcopyright{cagov}
%\setcopyright{cagovmixed}

% DOI
%\doi{http://dx.doi.org/}

% ISBN
%\isbn{978-1-4503-4073-1/16/10}

%\acmPrice{\$15.00}

\clubpenalty=10000 
\widowpenalty = 10000

\title{Leveraging Social Signal to Improve Item Recommendation  for Matrix Factorization}

\numberofauthors{2}

\author{
Ze Wang, and Hong Li\\
{Dept. of Computer Science, Shandong University of Technology}\\
{Shandong, China}\\
{zwang02934@gmail.com, hli0329@gmail.com}
}

\maketitle

\begin{abstract}
Although Recommender Systems have been comprehensively studied in the past decade both in 
industry and academia, most of current recommender systems suffer from the following issues:
1) The data sparsity of the user-item matrix seriously affect the recommender system quality. 
As a result, most of traditional recommender system approaches are not able to deal with
the users who have rated few items, which is known as cold start problem in recommender system.
2) Traditional recommender systems assume that users are independently and identically distributed
and ignore the social relation between users. However, in real life scenario, due to the exponential
growth of social networking service, such as facebook and Twitter, social connections between different 
users play an significant role for recommender system task. In this work, aiming at providing a better
recommender systems by incorporating user social network information, we propose a matrix factorization
framework with user social connection constraints. Experimental results on the real-life dataset
shows that the proposed method performs significantly better than the state-of-the-art approaches in terms of
MAE and RMSE, especially for the cold start users. 
\end{abstract}

\section{Introduction and Related Work}

As the exponential growth of information generated from the website, the recommender systems
have become more and more popular in both academia and industry. Recommender systems are 
a subclass of information filtering techniques that seek to predict the rating or preference
that user would give to an item (movies, books, music, restaurant).

Although recommender systems have been widely used, such as Google, Amazon, Ebay, most of 
recommender systems suffer from several common drawbacks. First of all, data sparsity is 
the most serious problems for designing effective and efficient recommender systems. As 
indicated in ~\cite{Sarwar.Karypis.ea:01,Zhang.Noman.ea:17,zhang2017purdue}, the density of data is less than $1\%$ in most of recommender systems. In 
other words, in typical user-item rating matrix, the number of nonzero ratings in this matrix
is less than $1\%$. This phenomenon is obvious and for example, customers in real life always purchase
very few items in amazon compared to the whole items in amazon database. So most of collaborative
filtering based recommender systems such as ~\cite{Xia.George:12,Breese.John.ea:98,Canny.John:02,Deshpande.Mukund.ea:04,Herlocker.John.ea:99,Hao.Irwin.ea:07,Santosh.Xia.ea:13,zhang2016trust,tkde_online} cannot handle the users who have rated few items. What is more, traditional recommender systems ignore the social connection among users in both social based network and
trust based network. It assumes that all the users are independently and identically distributed and only 
use the user-item matrix for recommendation task. However, in real life, we always turn to our trusted
friend for suggestions for buying books, cloth, and watching movies, etc. So our preference can be easily
affected by our friends. Therefore, simply ignoring social connection is not realistic when doing recommendation
systems and considering the social connection information in recommender system can efficiently deal with
the cold start users~\cite{Zhang.Jie.ea:14,icmla-dundar}. 

In this paper, aiming at solving the above problems, we propose a matrix factorization framework with
social network information as regularization term. The proposed model effectively handles the information
from two resources, user-item matrix and user social/trust network. More specifically, the social network 
information is used for designing the social regularization term to constrain the matrix factorization
objective function. The experimental analysis on one real-life dataset shows that our proposed model
outperforms several state-of-the-art algorithms.

Social based recommendation system has been studied in~\cite{Ma.Zhou.ea:11,Hao.Irwin.ea:09,Jamali.Martin.ea:10}. Furthermore, matrix factorization is widely used in the areas of social network analysis~\cite{icde-sutanay,Dave2018,chen2016incremental,Chen2017}, and name entity disambiguation~\cite{Zhang.Hasan.ea:15,Zhang.Saha.ea:14,Zhang.Dundar.ea:16,Zhang.Hasan.3}.
The rest of this paper is organized as follows: The problem definition and preliminary background are presented
in section 2. Basic matrix factorization framework is introduced in section 3. The proposed model is described 
in section 4. Our experiment is reported in section 5. Finally I conclude the paper.

\section{Problem Definition and Preliminary Background}

In recommender system, we have a set of users $U = \{u_1, u_2, ... u_M \}$, and a set of items
$I = \{i_1, i_2, ... i_N \}$. The user-item rating matrix is represented as $R = [{R_u}_i]_{M \times N}$
and ${R_u}_i$ is denoted as the rating of user $u$ on item $i$. Typically the domain of rating is the real 
number ranging from $1$ to $5$. 

For social network, given a directed network $G = (V, E)$, wehre every user $u \in V$, $(u, v) \in E$
if user $v$ is a directed neighbor of user $u$, in other words $v \in N_u$ and $N_u$ is a set of directed
neighbors of user $u$. For example, in trust network, if user $u$ likes the item review that user $v$ writes,
then there is a outgoing edge from $u$ to $v$, but at the same time $v$ doesn't necessarily link to $u$. 
Without losing generality, we denote the adjacent matrix of network $G$ as $ A = [{A_u}_v]_{M \times M}$. In this project,
the edge weight $w$ in matrix $A$ is a binary value $ \{0,1 \} $. For instance, if user $u$ has an outgoing edge to user $v$,
then ${w_u}_v = 1$, otherwise ${w_u}_v = 0$.

Assume that a given user $u \in U$, an item $i \in I$, the recommender task in this project is to predict the missing value 
${R_u}_i$ given matrix $R$ and $A$. In this work I utilize the matrix factorization based model to learn the latent factors
of users and items and predict the missing rating value in matrix $R$. 

\section{Basic Matrix Factorization Model}
In recommender system, an efficient approach for predicting missing values in user-item matrix is to employ 
matrix factorization method. Basically matrix factorization method is to factorize the user-item matrix and
use the low ranked user and item factor matrices for missing rating value prediction. 

Suppose given user-item matrix $R$, the basic matrix factorization model is as follows:

\begin{equation}
R \approx P^{T}Q
\end{equation}
where $R$ is the $M \times N$ user-item rating matrix, $P$ is the $K \times M$ user factor matrix and
$Q$ is the $K \times N$ item factor matrix, and $K$ is the number of latent space dimension for users
and items in user-item matrix $R$. 

The approximation of user $u$'s rating on item $i$, which is denoted by ${r_u}_i$, is defined as
$ {r_u}_i = p_u^{T}q_i $, where $p_u$ is the $K \times 1$ user factor for user $u$, $q_i$ is the
$K \times 1$ item factor for item $i$. And the objective function of matrix factorization model is 
as follows:

\begin{equation}\label{MF}
f(P, Q) = \frac{1}{2} \sum_{(u,i) \in R} {\parallel {r_u}_i - p_u^{T}q_i \parallel}^{2} + \frac{{\lambda}}{2}{{\parallel P \parallel}_{F}^{2}+\frac{{\lambda}}{2}{\parallel Q \parallel}_{F}^{2}}
\end{equation}
where $ \lambda > 0 $. We are going to minimize the $f$ so as to solve both $P$ and $Q$. In order to solve the 
optimization problem in equation ~\ref{MF}, a gradient descent approach can be used to obtain a local minimum.

\section{Social Connection Based Matrix Factorization Model}
Traditional recommender systems, like collaborative filtering, only utilize the user-item rating matrix information
for recommendation but ignore the social connections among users. Due to the factor that the online social network
is becoming more and more popular, incorporating social network information in recommender system becomes more and
more important. Particularly by embedding the social connection information in recommender system, it can efficiently
solve the cold start problem. In this section, I will incorporate the user social connection information as regularization
term into basic matrix factorization framework and use the gradient descent approach to solve the proposed model. 

\subsection{Social Regularization based Model}
We formulate the objective function as the following minimization problem:

\begin{equation}\label{optimize}
%\begin{multline*}
\begin{split}
\underset{P,Q}{\text{min}} F(R, P, Q) = \frac{1}{2} \sum_{(u,i) \in R} {\parallel {R_u}_i - P_u^{T}Q_i \parallel}^{2} + \frac{{\alpha}}{2} \sum_{u=1}^{m}\sum_{f \in F^{+}(u)}{Sim(u,f)\parallel P_u - P_f \parallel}^{2}\\
+ \frac{{\lambda}}{2}{{\parallel P \parallel}_{F}^{2}+\frac{{\lambda}}{2}{\parallel Q \parallel}_{F}^{2}}
\end{split}
%\end{multline*}
\end{equation}

In the formulation above, $P_u$ is the $k \times 1$ user factor for user $u$
(the u-th column in $P$), $Q_i$ is the $k \times 1$ item
factor for item $i$ (the i-th column in $Q$).
$Sim(u,f) \in \left[0, 1\right]$ is the similarity
function to indicate the similarity
between user $u$ and user $f$. Also we use $F^{+}(u)$ to denote user $u$'s outlink
friends and use the notation $F^{-}(u)$ to represent user $u$'s inlink friends. In
trust network, like Epinion, $F^{+}(u)$ doesn't necessarily equal to $F^{-}(u)$ since
trust network is directed network typically.

A local minimum of the objective function given by the objective function can be found
by performing the gradient descent with respect to latent vectors $P_u$ and $Q_i$.
The derivation procedure for user and item latent factors are as follows:

\begin{align*}
\begin{split}
%\begin{multline*}
\frac{\partial F}{\partial P_u} ={}& \sum_{i \mid (u,i) \in R} {(P_u^{T}Q_i - {R_u}_i) Q_i}+{\lambda P_u} 
+ \alpha \sum_{f \in F^{+}(u)}{Sim(u,f)\parallel P_u - P_f \parallel} \\  
                                   &+\alpha \sum_{g \in F^{-}(u)}{Sim(u,g)\parallel P_u - P_g \parallel} 
\end{split}\\
%\end{multline*}
%\begin{equation*}
%\begin{multline*}
\frac{\partial F}{\partial Q_i} ={}& \sum_{u \mid (u,i) \in R} {(P_u^{T}Q_i - {R_u}_i)P_u} + \lambda Q_i
%\end{multline*}
%\end{equation*}
\end{align*}

One of the advantages to employ this model is that the proposed model can capture the trust propagation of different
users. For instance, if user $u$ has a outgoing friend $v$, and $v$ also has a outgoing friend $g$, but at the same time
$u$ and $g$ are not friend based on the social network connection. Using this model, it can indirectly minimize the feature 
distance between $u$ and $g$. 

\subsection{Similarity Measurement}
in the proposed model, we need to quantify the similarity between users in social network.
Given the knowledge of user-item matrix, we can model the user user similarity based on purchased items and 
corresponding ratings. In order to achieve this goal, Pearson Correlation Coefficient (PCC) and Vector Space Similarity (VSS) are proposed
to define similarity between different users. The equations are as follows:

\begin{align*}
%\begin{equation*}
PCC(i,f) &= \frac{\displaystyle \sum_{j \in I(i) \cap I(f)}({R_i}_j- \overline{R_i})({R_f}_j- \overline{R_f})}{\sqrt{\sum_{j \in I(i) \cap I(f)}({R_i}_j- \overline{R_i})^{2}} \cdot \sqrt{\sum_{j \in I(i) \cap I(f)}({R_f}_j- \overline{R_f})^{2}}} \\
%\end{equation*}
%\begin{equation*}
VSS(i,f) &= \frac{\displaystyle \sum_{j \in I(i) \cap I(f)}{R_i}_j \cdot {R_f}_j}{\sqrt{\sum_{j \in I(i) \cap I(f)}{R_i}_j^{2}} \cdot \sqrt{\sum_{j \in I(i) \cap I(f)} {R_f}_j^{2}}}
%\end{equation*}
\end{align*}
where $\overline{R_i}$ represents the average rate of user $i$ and $j$
belongs to the subset of items which user $i$ and user $f$ both rated.
${R_i}_j$ is the rate user $i$ on item $j$.

From the above definitions, we can see that $VSS(i,f) \in [0,1]$ and $PCC(i,f) \in [-1, 1]$. 
The larger values for both VSS and PCC, the more similar between different users. Also in order to
constrain the range of $PCC$ measurement into $[0,1]$, I use a simple mapping function $f(x) = \frac{x+1}{2}$ to bound its
similarity range from $0$ to $1$.

\section{Experiments and Results}

\subsection{Epinions Trust Dataset}
The dataset we employ for this work is Epinions\footnote{http://www.epinions.com}.
In Epinions, every user can read the reviews about a variety of items and also users can
write a review for particular items. For the social network viewpoint, 
each member in Epinions has a trust list of other members to indicate if I trust
your review or not. From my study, the dataset contains $49289$ different users
and $139738$ different items. And the total number of rating is $664824$. 

In order to verify the correctness of similarity measurement in my model, the first 
study is to validate my assumption that in trust network such
as Epinions, social friends have similar tastes. We utilize Vector Space
Similarity (VSS) as the metric to evaluate the similarity between user $i$
and user $j$.
 
The analysis we conduct is to understand how does the social friends similarity
compare with random peer similarity? The detailed analysis is as follows:

1. For each user $i$ in trust network, we calculate the average social friends' similarity
as follows:

\begin{equation}
\overline{S_i} = \frac{\sum_{k \in F^{+}(i)}{S_i}_k}{\mid F^{+}(i) \mid}
\end{equation}

where $F^{+}(i)$ is the trusted list of user $i$ which also means user $i$'s
outgoing friends.

2. For each user $i$, we also calculate the average random peer similarity as follows:

\begin{equation}
\overline{R_i} = \frac{\sum_{k \in R(i)}{S_i}_k}{\mid R(i) \mid}
\end{equation}

where $R(i)$ represents the random peer list of user $i$, which has same size with
$F^{+}(i)$ and $R(i) \cap F^{+}(i) = \emptyset$.

The motivation of carrying out this experiment is that we want to confirm that
the social peer relation in trust network has strong positive correlation
with user interest similarity. In order to reduce the noise of dataset, we only
consider the users who has more than five outgoing social peers in Epinions. For
those users who has less than five outgoing social peers, we simply ignore them 
in the evaluation.

In order to quantify the correlation between social relations and user interest 
similarity, we would like to measure the proportion of users whose social similarities
are greater than their random similarities followed by the equation 
$\overline{S_i} - \overline{R_i} > 0$. In the Epinions dataset, based on my study, there are
$73.4\%$ users whose social similarities are greater than their random similarities.
So we can safely make conclusion that friends have similar taste in Epinions.

We also test PCC based similarity measurement. We observe the similar result so for simplicity, 
therefore we only report the result using VSS similarity function.

\subsection{Evaluation Metrics}
The quality of the results is measured by the the Mean Absolute Error (MAE)
and the Root Mean Square Error (RMSE). The MAE and RMSE are defined as follows:

\begin{equation*}
MAE = \frac{1}{T} \sum_{i,j}{\mid {R_i}_j - \hat{{R_i}_j} \mid}
\end{equation*}

\begin{equation*}
RMSE = \sqrt{\frac{1}{T} \sum_{i,j}{({R_i}_j - \hat{{R_i}_j})^{2}}}
\end{equation*}

where ${R_i}_j$ denotes the rating user $i$ gave to item $j$,
$\hat{{R_i}_j}$ is represents the rating user $i$ gave to item
$j$ which is predicted by my proposed method, and $T$ is the total
number of nonzero ratings in the test dataset.

From the definition, we can see that the smaller MAE and RMSE, the better
performance of our proposed method.

\subsection{Comparison Methods}

Several of state-of-art methods are used for comparison with the proposed approach in this work. 
The detailed descriptions of comparison methods are explained below:

$\bullet$ UserMean: this method uses the mean values of every
user to predict the missing values

$\bullet$ ItemMean: the method utilizes the mean value of every item to
predict the missing values.

$\bullet$ BasicMF: Traditional Matrix Factorization and the method uses classic matrix
factorization formulation without incorporating user trust network information.

$\bullet$ Proposed Method: This method uses classic matrix factorization formulation with
trust network information as regularization term which is the method proposed in this work.

The results we obtain are shown in the tables ~\ref{tab:comparison90} and ~\ref{tab:comparison80}. In the whole process, we use PCC for user
user similarity measurement. During the experiment, parameters $\lambda$ is set to a trivial
value 3.0 and $\alpha$ in our model is set to 0.01. For the Epinions dataset,
we use 90\% and 80\% of the original data as the training data settings and the remaining as the test set. The random selection is carried out
$5$ times independently, and I report the average MAE and RMSE values in table ~\ref{tab:comparison90} and table ~\ref{tab:comparison80}.

As we can see our proposed method performs much better than baseline methods on Epinion dataset.
For instance, traditional matrix factorization method achieves $0.8641$ and $1.1071$ on MAE and 
RMSE under $90\%$ training setting respectively, whereas our proposed method achieves $0.8324$ 
and $1.0756$ on MAE and RMSE respectively. And the performance of UserMean and ItemMean methods 
are even worse than basic matrix factorization method. Cross-validation t-test shows that our proposed 
method is significantly better (p-value 0.0042 for MAE and 0.0036 for RMSE) than traditional matrix factorization. 
Similar results are obtained compared with other two baseline methods and also $80\%$ training setting.

\begin{table}[h!]
\centering
\begin{tabular}{|c|c|c|}
\hline
Method & MAE  & RMSE  \tabularnewline
\hline
UserMean & 0.9415 & 1.2361 \tabularnewline
\hline
ItemMean & 1.2236 & 1.7985 \tabularnewline
\hline
MF & 0.8641 & 1.1071 \tabularnewline
\hline
Proposed Method & 0.8324 & 1.0756 \tabularnewline
\hline
\end{tabular}
\caption{Performance Comparisons (Dimensionality = 10 and 90\% training setting)}
\label{tab:comparison90}
\vspace{-0.2in}
\end{table}

\begin{table}[h!]
\centering
\begin{tabular}{|c|c|c|}
\hline
Method & MAE  & RMSE  \tabularnewline
\hline
UserMean & 0.9545 & 1.2489 \tabularnewline
\hline
ItemMean & 1.2663 & 1.8575 \tabularnewline
\hline
MF & 0.8722 & 1.1254 \tabularnewline
\hline
Proposed Method & 0.8462 & 1.0964 \tabularnewline
\hline
\end{tabular}
\caption{Performance Comparisons (Dimensionality = 10 and 80\% training setting)}
\label{tab:comparison80}
\vspace{-0.2in}
\end{table}

\subsection{Study of Parameter Sensitivity}
\begin{figure}[h]
\centering
\subfigure[Epinions (MAE)]
{
\label{fig:MAE}
\includegraphics[height=0.45\linewidth ,angle=-90] {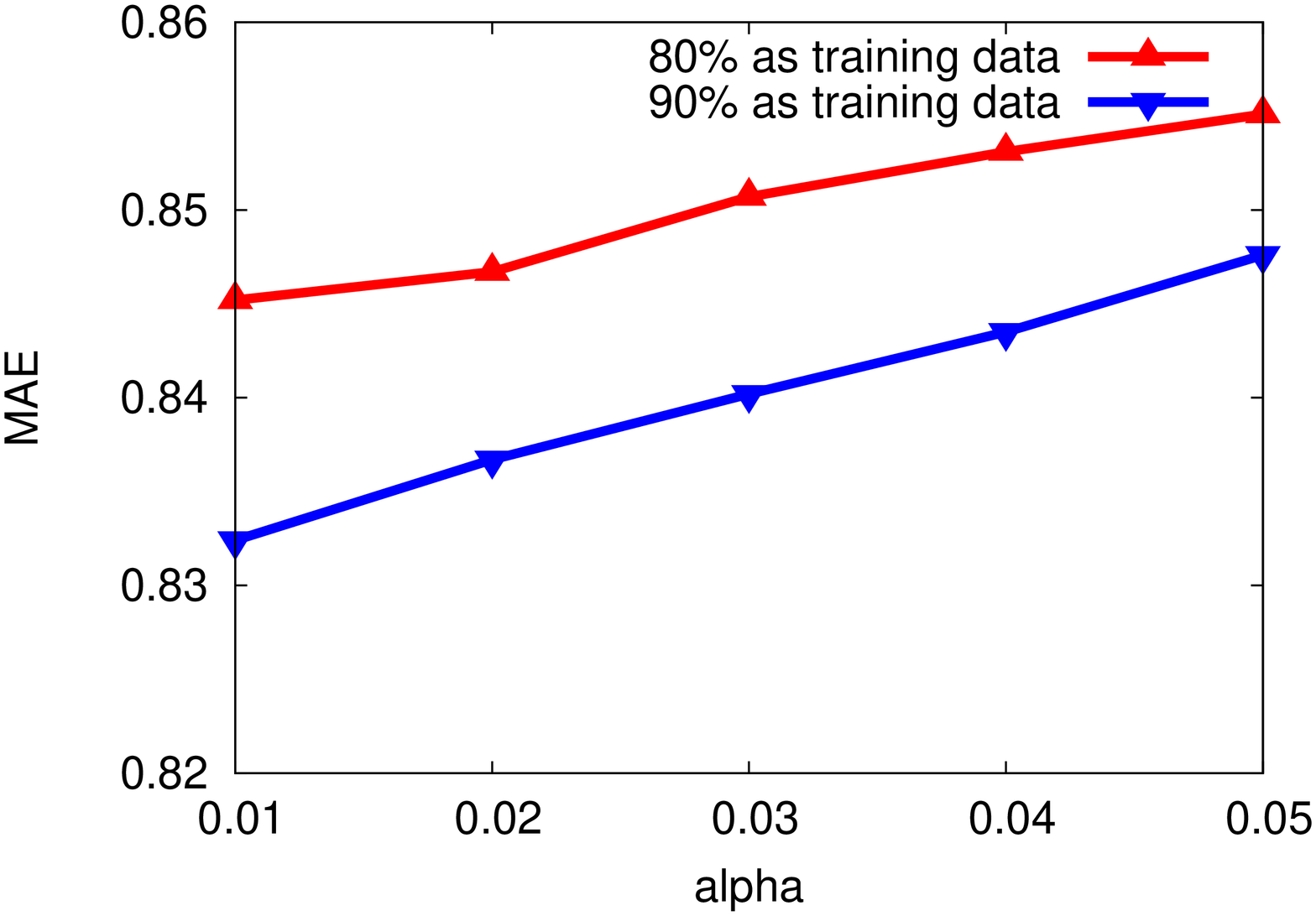}
}
\subfigure[Epinions (RMSE)]
{
\label{fig:RMSE}
\includegraphics[height=0.45\linewidth ,angle=-90] {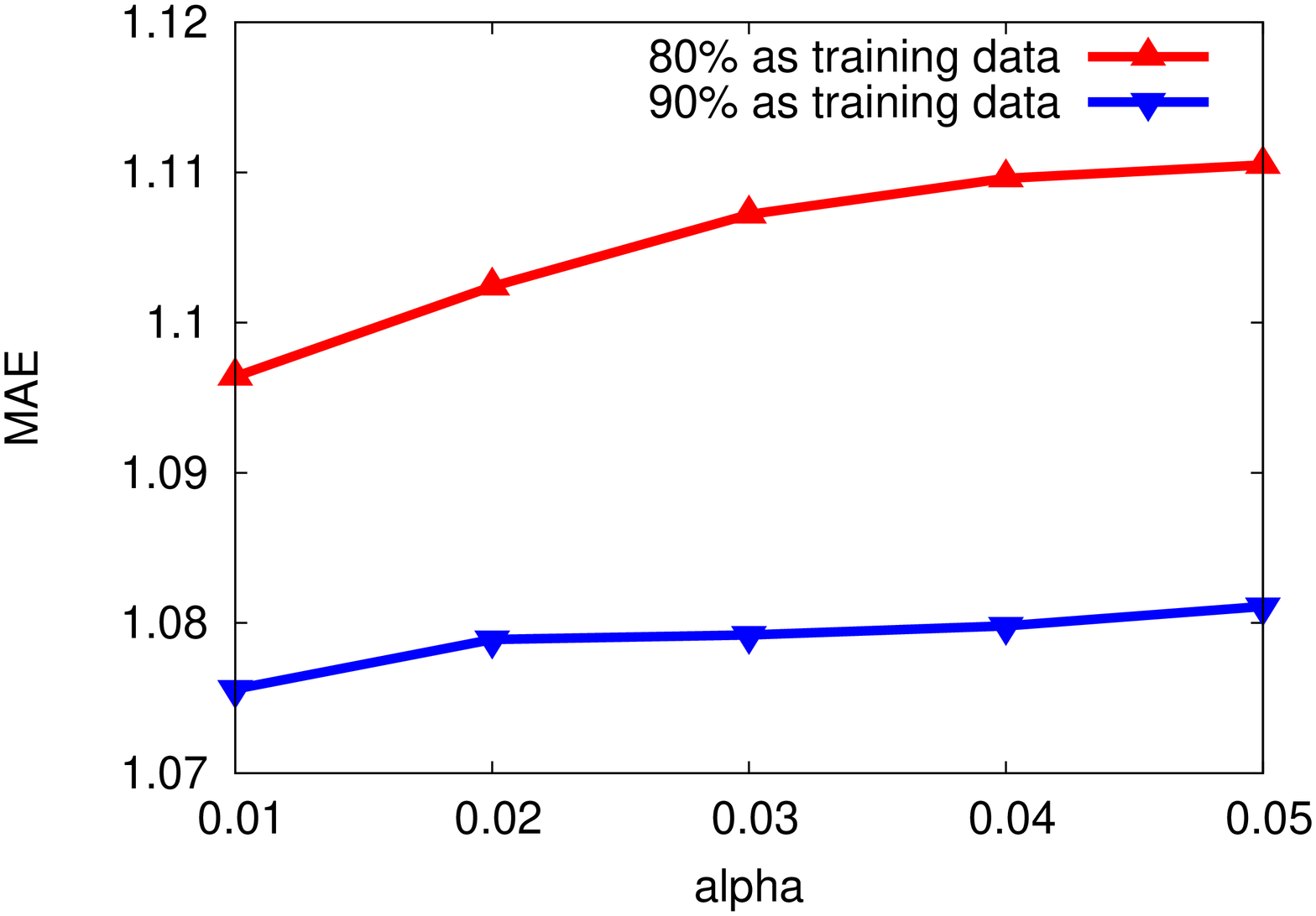}
}
\caption{Parameter Sensitivity of $\alpha$ with Dimensionality = 10}
\vspace{-0.2in}
\end{figure}

In our proposed method, the parameter $\alpha$ plays the important role for how much
our method should incorporate the social network information. Large value of $\alpha$
indicates more impact of social connection information in the propsoed model and smaller
value of $\alpha$ indicates less impact of social connection information and zero $\alpha$ value makes the proposed
model close to basic matrix factorizaton method. In this experiment, we can see how the performance of model
changes as we vary the value of $\alpha$ parameter. The result of this experiment is shown in Figure ~\ref{fig:MAE}
and ~\ref{fig:RMSE}. As we can see that the performance in terms of both MAE and RMSE degrades for the choice of $\alpha$.

\subsection{Impact of Similarity Function}
The similarity function $Sim(i,f)$ measures how similar of user $u_i$ and $u_f$
in the social network. In the proposed method, two similarity functions PCC and VSS
are used and in this section, in order to measure how similarity functions contribute
the proposed model, besides PCC and VSS, two other similarity frameworks will be
used for comparison:

1. Setting all the similarities between users as 1.

2. Assigning a random similarity between $0$ and $1$ to any pair of friendship.

The result is shown in the following tables ~\ref{tab:simi90} and ~\ref{tab:simi80}.

\begin{table}[h!]
\centering
\begin{tabular}{|c|c|c|}
\hline
Similarity & MAE  & RMSE  \tabularnewline
\hline
Sim=1 & 0.8415 & 1.1061 \tabularnewline
\hline
Sim=Random & 0.8636 & 1.1085 \tabularnewline
\hline
Sim=VSS & 0.8351 & 1.0789 \tabularnewline
\hline
Sim=PCC & 0.8324 & 1.0756 \tabularnewline
\hline
\end{tabular}
\caption{Similarity Analysis (Dimensionality = 10 and 90\% training setting)}
\label{tab:simi90}
\vspace{-0.2in}
\end{table}

\begin{table}[h!]
\centering
\begin{tabular}{|c|c|c|}
\hline
Similarity & MAE  & RMSE  \tabularnewline
\hline
Sim=1 & 0.8515 & 1.1121 \tabularnewline
\hline
Sim=Random & 0.8536 & 1.1145 \tabularnewline
\hline
Sim=VSS & 0.8491 & 1.0989 \tabularnewline
\hline
Sim=PCC & 0.8462 & 1.0964 \tabularnewline
\hline
\end{tabular}
\caption{Similarity Analysis (Dimensionality = 10 and 80\% training setting)}
\label{tab:simi80}
\vspace{-0.2in}
\end{table}

As we can see from the result that the performance of unweighted similarity 
and random similarity measurements are worse than VSS and PCC. Among these four
similarity functions, PCC performs best. So this observation demonstrates that
the similarity function plays important role in my proposed model.

\subsection{Performance on cold start users}
In this section we study how performance for our proposed model for the cold start users in recommender 
systems. Cold start users are users who have rated or purchased few items and traditional collaborative
filtering methods are not able to deal with this kind of users since cold start users don't have enough
rating/purchase history. In this experiment, we consider users who have rated less than 5 items as cold start users.
In Epinions, more than 55\% of users are cold start users. So proposing effective recommender systems for coping with
cold start users are becoming more and more important. For conducting this experiment, we only test on the cold start 
users and put one of their rated items into test set and other rated items into training set. The result is shown in table
~\ref{tab:cold-start}.

\begin{table}[h!]
\centering
\begin{tabular}{|c|c|c|}
\hline
Method & MAE  & RMSE  \tabularnewline
\hline
UserMean & 1.071 & 1.502 \tabularnewline
\hline
ItemMean & 1.082 & 1.582 \tabularnewline
\hline
MF & 0.982 & 1.261 \tabularnewline
\hline
Proposed Method & 0.892 & 1.121 \tabularnewline
\hline
\end{tabular}
\caption{Performance on cold start users (Dimensionality = 10 and $\alpha = 0.01$)}
\label{tab:cold-start}
\vspace{-0.2in}
\end{table}

As we can see from the result, our proposed method could deal with the cold start users and the 
performance is much better than other three baseline methods.

\section{Conclusion}
In this paper, we leverage the social network information in the matrix factorization framework
to obtain a better recommender system, especially for the cold start users. The experimental 
analysis on the real-life dataset shows that our proposed approach performs significantly
better than several state-of-the-art methods.

\medskip

\balance

%\bibliography{sample}lance
\bibliographystyle{abbrv}
\bibliography{final_report}  % sigproc.bib is the name of the Bibliography in this case

\end{document}